\documentclass[useAMS,usenatbib,usegraphicx]{mn2e}
\usepackage{natbib}
\usepackage{times}

\def\ltsima{$\; \buildrel < \over \sim \;$}
\def\simlt{\lower.5ex\hbox{\ltsima}}
\def\gtsima{$\; \buildrel > \over \sim \;$}
\def\simgt{\lower.5ex\hbox{\gtsima}}

\title[Mergers in dwarf spheroidals]{Mergers and the outside-in formation of dwarf spheroidals}
\author[A. Ben\'itez Llambay et al.]{A. Ben\'itez-Llambay$^{1,2}$\thanks{E-mail:
alejandrobll@oac.uncor.edu}, J. F. Navarro$^{3}$, M. G. Abadi$^{1,2}$, S. Gottl\"ober$^{4}$,
\& \newauthor G. Yepes$^{5}$, Y. Hoffman$^{6}$, M. Steinmetz $^{4}$\\
$^{1}$Observatorio Astron\'omico, Universidad Nacional de C\'ordoba, C\'ordoba, X5000BGR, Argentina\\
$^{2}$Instituto de Astronom\'ia Te\'orica y Experimental - CONICET, Laprida 922 X5000BGR, C\'ordoba, Argentina\\
$^{3}$Senior CIfAR Fellow. Department of Physics \& Astronomy, University of Victoria, Victoria, BC, V8P 5C2, Canada\\
$^{4}$Leibniz Institute for Astrophysics, An der Sternwarte 16, 14482 Potsdam, Germany\\
$^{5}$Departamento de F\'isica Te\'orica, Universidad Aut\'onoma de Madrid, 28049, Madrid, Spain\\
$^{6}$Racah Institute of Physics, The Hebrew University of Jerusalem, Jerusalem, 91904, Israel}

\begin{document}

\date{}

%\pagerange{\pageref{firstpage}--\pageref{lastpage}} \pubyear{2002}

\maketitle

\label{firstpage}

\begin{abstract}
  We use a cosmological simulation of the formation of the Local Group to explore the origin of age and metallicity gradients in dwarf spheroidal galaxies. We find that a number of simulated dwarfs form ``outside-in'', with an old, metal-poor population that surrounds a younger, more concentrated metal-rich component, reminiscent of dwarf spheroidals like Sculptor or Sextans. We focus on a few examples where stars form in two populations distinct in age in order to elucidate the origin of these gradients. The spatial distributions of the two components reflect their diverse origin; the old stellar component is assembled through mergers, but the young population forms largely in situ. The older component results from a first episode of star formation that begins early but is quickly shut off by the combined effects of stellar feedback and reionization. The younger component forms when a late accretion event adds gas and reignites star formation. The effect of mergers is to disperse 
the old stellar population, increasing their radius and decreasing their central density relative to the young population. We argue that dwarf-dwarf mergers offer a plausible scenario for the formation of systems with multiple distinct populations and, more generally, for the origin of age and metallicity gradients in dwarf spheroidals.  \end{abstract}


\section{Introduction}
\label{SecIntro}

Dwarf spheroidal galaxies (dSphs) are low-mass, gas-free, dark matter-dominated systems whose origin is still uncertain. They are not currently forming stars and show prominent old stellar populations, but their star formation histories (SFHs) are not uniformly old \citep[see the recent reviews of][]{Tolstoy2009,Weisz2014}. Indeed, dSphs show a wide variety of SFHs; from systems that essentially completed forming their stars $\sim 10$ Gyr ago, like Cetus and Draco~\citep{Monelli2010,Aparicio2001} to others where star formation was sustained over a long time and ceased only as recently as $6$ Gyr ago~\citep[e.g., Leo II, ][]{Dolphin2002} or even $<1$ Gyr ago~\citep[e.g., Fornax,][and references therein]{Stetson1998, Saviane2000, Battaglia2006, deBoer2012}.

Indeed, the structural properties of dSphs \citep[i.e., density profile shape, half-light radius, velocity dispersion, central surface brightness, overall metallicity; see, e.g.,][and references therein]{Kormendy2012,Kirby2013} are similar to those of other dwarf galaxies of similar stellar mass but where star formation is still ongoing today or ceased very recently, and whose optical morphology, dominated by patches of young stars, is more irregular (i.e., dwarf ``irregular'' galaxies, dIrrs, and dwarf ``transition'' systems, dTrans). This has led some to argue that there is no substantial difference between morphological types; in this view, dIrrs and dTrans would just be dwarf spheroidals with some residual star formation at recent times.

One problem with this interpretation is that the kinematics of stars differs between morphological types. Whereas there is little evidence for systemic rotation in dSph galaxies, a number of dIrr (or at least their gas and young stars) show clear signs of ordered rotation, at least in their gas component and young stars \citep[e.g.,][]{Leaman2012,Kirby2014}. Although finding rotating subcomponents in a dSph would be harder than in dIrr, where recently-formed stars are more easily identified, it is generally agreed that the kinematic evidence suggests a scenario where some event in the past history of dSphs has been responsible for both terminating their star formation and for reducing the importance of centrifugal support \citep[see, however,][for a more nuanced view]{Wheeler2015}.

Mergers are an obvious possibility \citep[see, e.g.,][]{Mihos1996,Kazantzidis2011,Helmi2012,Starkenburg2015}, but a compelling case has also been made for the repeated influence of Galactic tides, a process usually termed ``tidal stirring'' \citep{Mayer2001,Read2006,Mayer2010}.  The latter, in particular, explains naturally why most dSphs in the Local Group (where they are best studied given their low luminosity) are at present satellites of the Milky Way or Andromeda galaxies \citep{vandenBergh1994,Grebel1999,Mateo2008}. Tidal stirring, however, fails to explain the origin of {\it isolated} dSphs such as Cetus (or Tucana), whose large distance from any major galaxy excludes repeated tides as the main mechanism driving its morphology, or its transformation. Tidal stirring also faces difficulties explaining the origin of extremely dark matter-dominated dSphs, whose progenitors would have been much less prone to develop the bar and buckling instabilities responsible, in that scenario, for transforming disks 
into spheroids.

Distinguishing between a tidal or a merger origin of {\it satellite} dSphs is more problematic, since both mechanisms share a number of similar features. Tidal stirring deforms galaxies, reduces rotation, generates tidal tails, and brings gas to the centre, where it may be consumed in a burst of star formation, just like mergers would. Discerning between merging or stirring thus requires considering other properties beyond rotational support or evidence for past tidal interaction. 

One possibility is to consider the time when star formation ceased in various satellite dSphs; if stirring dominates, this should correlate closely with the time of first infall into the Galaxy's halo, which could, in principle, be inferred from the orbital energy of the satellite \citep[see, e.g.,][]{Rocha2012}. However, the possibility that star formation may be brought to an end by effects other than Galactic tides, such as cosmic reionization \citep{Bullock2000,Benson2002,Susa2004,Ricotti2005} or ``cosmic web stripping'' \citep{Benitez-Llambay2013}, hinders in practice the applicability of this idea.

The presence, at least in some dSphs, of radial gradients in age and metallicity provides an important clue. When present, these gradients typically indicate that the younger, metal-rich component is surrounded by a more extended envelope of older, metal-poor stars. This ``outside-in'' formation process is the reverse of prevailing gradients in disk galaxies, where younger stars populate preferentially the outskirts of a galaxy. These gradients are particularly clear in some galaxies and, when accompanied by correlated differences in kinematics and metallicity, are often interpreted as evidence for distinct stellar components. The Sculptor dSph makes a compelling example of a system with at least two distinct components \citep{Tolstoy2004,Battaglia2008}, but so do other dwarfs such as Sextans \citep{Battaglia2011}; the And II companion to M31 \citep{McConnachie2007,Amorisco2014}; and Fornax \citep{Battaglia2006,Walker2011}. We note, however, that clear-cut evidence for distinct components is often weak, and that 
these may also be interpreted as simply cases where the radial gradients are particularly pronounced.

The presence of such components, especially when clearly distinct in metallicity/age and kinematics/spatial distribution, would seem to favour a merger scenario, since the sudden accretion of external material could then easily explain how separate components with distinct stellar populations may arise in the same galaxy. 

Mergers, however, are not the only possible formation path for age gradients or multiple components. \citet{Schroyen2011,Schroyen2013} cite rotation (or lack thereof) as a key ingredient, while episodic star formation has also been suggested~\citep[see the discussion by][]{Tolstoy2004,Battaglia2006}; a scenario where feedback from the evolving stars of one star formation episode pushes gas out of the galaxy, prompting a hiatus in star formation until the gas can return and ignite a further episode later on. This, however, fails to explain the origin of the spatial segregation between components. Why would the returning gas lead to the formation of a more concentrated stellar component?

\begin{figure*}
 \begin{center}
 \includegraphics[scale=0.36]{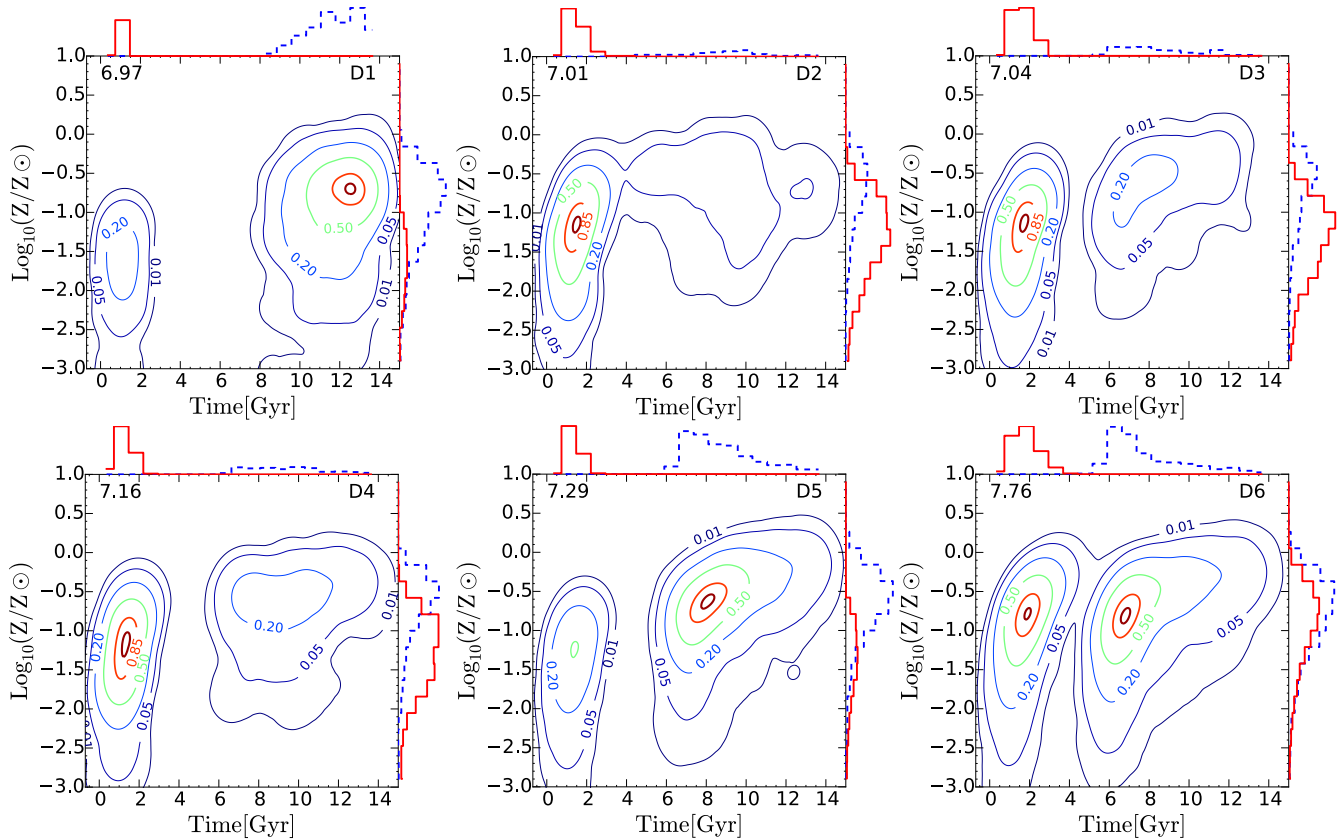}
 \caption{Formation time-metallicity distribution of stars in the six ``two-component'' simulated dwarfs at redshift $z=0$. Contours indicate the number density of stars in the age-metallicity plane, and they are labelled according to the value of the density respect to its maximum. Histograms on top and right of each panel indicate the distribution of stars along the formation time and metallicity axes, respectively. Red (solid) is used to indicate the ``old'' component; blue (dashed) for the ``young'' component. Note that each of these systems were chosen to have two components of distinct age. Each of these components, however, seems to also have different metallicities. The numbers in each panel label each dwarf for subsequent reference (``D1'' to ``D6''). The total stellar mass of each dwarf is also listed (in decimal log units relative to the Sun).} \label{FigAgeMet}
 \end{center}
\end{figure*}

\begin{figure*}
 \begin{center}
 \includegraphics[scale=0.5]{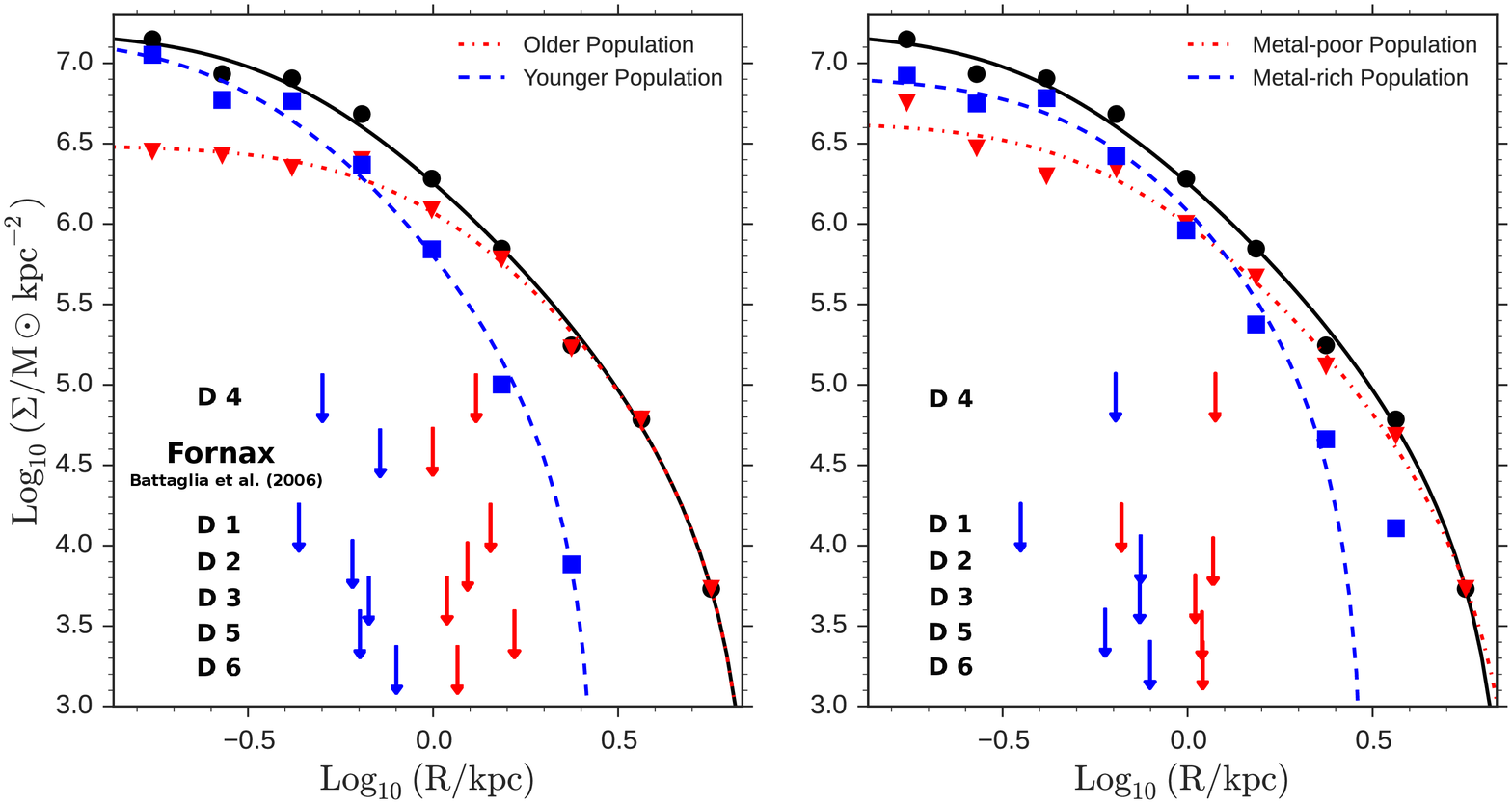}
 \caption{Projected stellar density profile of the simulated dwarf ``D4'' (black circles). In the left panel, inverted triangles and solid squares correspond to the ``old'' and ``young'' stellar populations (i.e., $t_{\rm form}$ smaller or greater than $4.5$ Gyr, respectively, see Fig.~\ref{FigAgeMet}). The right panel shows the same, but the components are split by metallicity, according to what stars are more or less metal rich than $[Z/Z_\odot]=-1$. Curves indicate \citet{King1962} model fits to each profile.  Top arrows indicate the projected half-mass radius of each component; the rest of the arrows show the same, but for the other simulated dwarfs (profiles not shown). The arrows labelled ``Fornax'' correspond to the half-number radii of the intermediate-age and older populations derived by \citet{Battaglia2006}. Note that the old/metal-poor component is always less centrally concentrated than their young/metal-rich counterpart.}
 \label{FigMassProf}
 \end{center}
\end{figure*}
%%%%%%%%

An alternative scenario invokes selective heating by external ionizing radiation, which would shut off star formation in the outer (less dense) regions first, allowing better-shielded (denser) gas near the centre to continue forming stars for an extended period of time, self-enriching subsequent populations of stars and leaving behind a system with a strong metallicity and age gradient \citep{Kawata2006}. This scenario, however, fails to explain in a simple way why the components, in galaxies like Sculptor at least, should differentiate so neatly.

The outside-in formation of dwarf spheroidals might therefore be more easily explained by external, rather than internal, mechanisms. It is in this context that we study the properties of simulated dwarf galaxies formed in a cosmological hydrodynamical simulation of the Local Group,  tailored to match the broad properties of the Milky Way and Andromeda galaxies as well as the surrounding large-scale structures.

As we discuss below, these simulated dwarfs provide insight into the outside-in formation process of dSphs, as well as into the origin of differences in their kinematic and enrichment properties. We begin with a brief description of the simulations (Sec.~\ref{SecSims}) and explore in Sec.~\ref{SecResults} the properties of the simulated dwarfs and their origin. We conclude with a brief summary of our main conclusions in Sec.~\ref{SecConc}.

\section{Numerical Simulations}
\label{SecSims}

\subsection{The CLUES simulation}

The simulation used in this work is part of the CLUES\footnote{Constrained Local UniversE Simulations: www.clues-project.org} project, which evolves realizations of the local universe, including the constraints placed by nearby large-scale mass distributions and the tidal fields that arise from them. These simulations have been presented by~\citet{Gottloeber2010} and several aspects of them have been reported and discussed in recent work~\citep[e.g.,][]{Libeskind2010,Knebe2011,DiCintio2012,Benitez-Llambay2013, Benitez-Llambay2014}. We  discuss briefly the main features relevant to our analysis below and refer the interested reader to those papers for details.

Table~\ref{TabSimParam} lists the parameters of the ``WMAP3'' simulation chosen for our analysis. This is a ``zoomed-in'' simulation of the Local Group where a small volume at the centre of the $64$ Mpc/h box of the main CLUES run has been resimulated at high resolution using the {\small GADGET-2} code. The simulation includes star formation, metal enrichment and supernova-driven winds (which can transport metals) using the multiphase model of~\cite{Springel2003}, as well as a cosmic ionizing UV background that follows the prescriptions of \cite{Haardt1996}.

\begin{figure*}
 \begin{center}
 \includegraphics[scale=0.36]{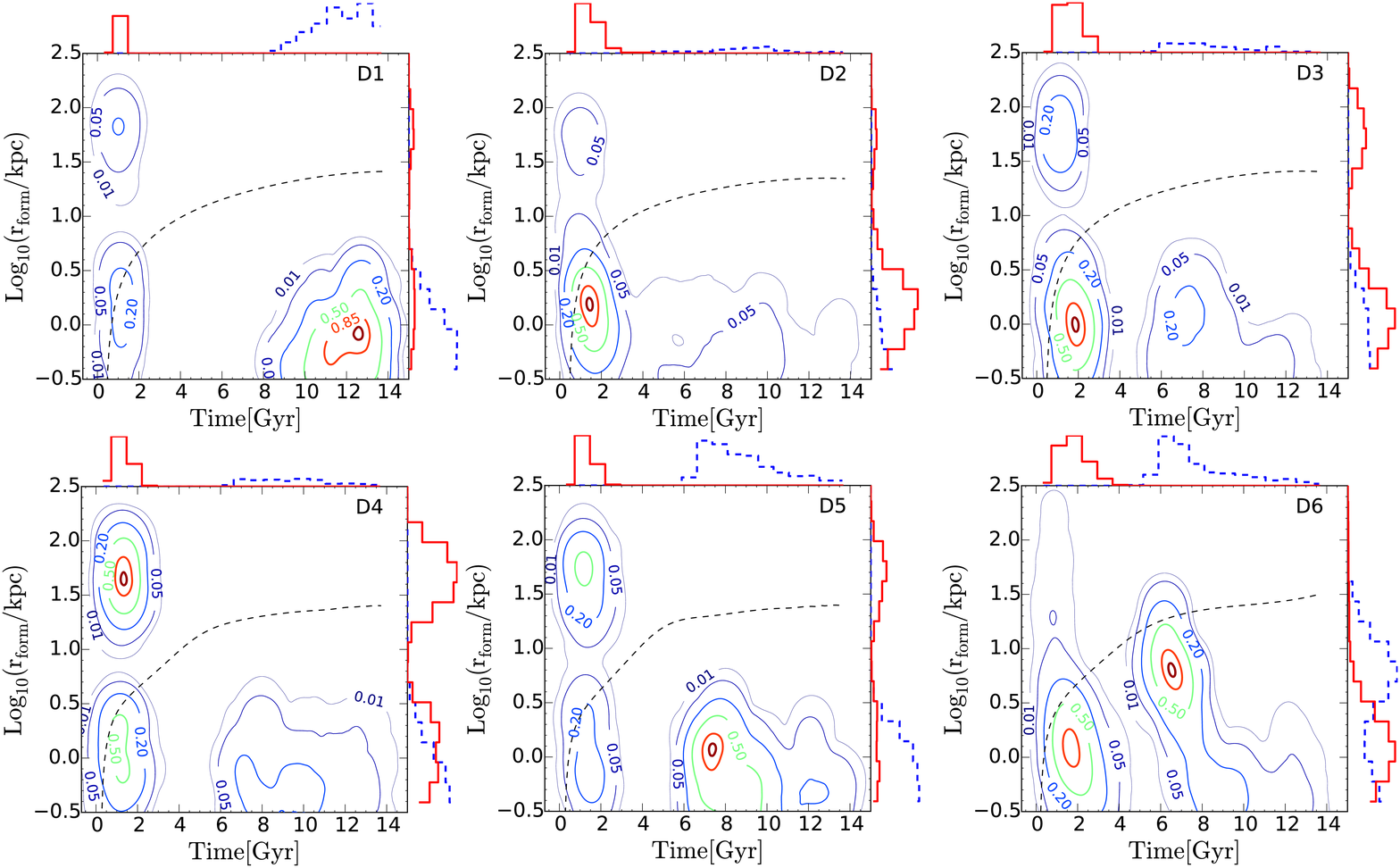}
 \caption{Distribution of formation times and of the distance to the centre of the most massive progenitor at the time of formation for all six simulated dwarfs. The format of the figure is analogous to Fig.~\ref{FigAgeMet}. Black dashed lines show the evolution of the virial radius of the most massive progenitor. Stars formed outside this line clearly belong to another progenitor that was later accreted into the main galaxy. The colour coding of the histograms indicates whether they correspond to the ``old'' (red solid) or ``young'' (blue dashed) stellar component.  Note that the distribution of distances of the older stellar population is typically bimodal, indicating that the older stellar component was assembled through mergers. The young stellar population, on the other hand, was formed largely in situ.} \label{FigAgeDist}
 \end{center}
\end{figure*}

\subsection{The simulated dwarfs}

We follow the same analysis procedure described in \citet{Benitez-Llambay2014}. In brief, we focus on a sample of dwarf galaxies identified in a $2$ Mpc-radius sphere centred at the barycentre of the two most massive galaxies (the Milky Way and M31 analogues) at $z=0$. 
%new paragraph
We have searched for dwarfs in this region using the group finder {\small SUBFIND} \citep{Springel2001}, applied to a list of friends-of-friends (FoF) halos constructed with a linking-length of $0.2$ times the mean interparticle separation.

  {\small SUBFIND} recursively identifies self-bound substructures in each halo, and provides a list of systems both ``central'' to each FoF halo as well as their ``satellites''. We will in what follows identify ``centrals'' with isolated dwarfs, after verifying that none of these has been, in the past, satellite of a more massive system. Our dwarf sample, therefore, contains both systems that have evolved in isolation in the periphery of the massive galaxies of the Local Group and satellites that have been affected by additional processes such as tidal stirring, ram-pressure stripping, etc.

  We include in the isolated dwarf sample only systems with virial\footnote{We define the virial mass, $M_{200}$, of a halo as that enclosed by a sphere of mean density $200$ times the critical density of the Universe, $\rho_{\rm crit}=3H^2/8\pi G$. Virial quantities are defined at that radius, and identified by a ``200'' subscript.} masses above $3\times 10^9 \, M_\odot$, which corresponds to roughly $\sim 10^4$ particles. For each system we measure the stellar mass, $M_{\rm gal}$, of each dwarf within a ``galactic radius'' defined as $r_{\rm gal}=0.15\, r_{200}$. We also restrict our analysis to systems with virial mass below $10^{10}\, M_\odot$ so that our analysis focuses on systems that span the range $6.3 \times 10^6 \simlt < M_{\rm gal}/M_\odot \simlt 6.3 \times 10^7$. We use the latter criterion as well in order  to select satellites retained for analysis.

We characterize the SFH of each dwarf by computing the fraction of stars formed in three different intervals $\Delta t$ of cosmic time: $f_{\rm old}$ corresponds to ``old'' stars formed in the first $4$ Gyr of cosmic evolution; $f_{\rm int}$ to ``intermediate-age'' stars ($4<t_{\rm form}/$Gyr$<8$), and $f_{\rm young}$ to stars formed after the Universe was $8$ Gyr old.
We express these fractions as star formation rates (SFR) normalized to the past average, ${\bar f}=M_{\rm gal}/t_0$, where $t_0=13.7$ Gyr is the age of the Universe. In other words, 
\begin{equation}
f_j={1\over X}{M_ j/\Delta t_j \over {\bar f}}, 
\end{equation}
where the subscript $j$ stands for either the ``old'', ``intermediate'', or ``young'' component, and
\begin{equation}
X={1 \over {\bar f}}\sum_j M_j/\Delta t_j
\end{equation}
is a normalizing coefficient that ensures that $f_{\rm old}+f_{\rm int}+f_{\rm young}=1$. A galaxy that has formed stars at constant rate thus has $f_{\rm old}=f_{\rm int}=f_{\rm young}=1/3$.

%%new paragraph
We use this parametrization to select systems with unusual SFHs, where radial gradients in age and metallicity might be expected. As mentioned in Sec.~\ref{SecIntro}, examples of such systems include dwarfs with distinct stellar populations. We therefore focus our analysis on those that, according to their SFH, host at least two populations of distinct age. In particular, we retain systems whose SFHs satisfy the following three criteria\footnote{These are the six systems identified by solid cyan triangles in Figs.~6 and 7 of \citet{Benitez-Llambay2014}}: $f_{\rm int}<0.2$; $f_{\rm old} > 0.1$; and $f_{\rm young} > 0.1$. These criteria select systems that have experienced a hiatus in their star formation activity at intermediate times (the upper limit imposed on $f_{\rm int}<0.2$) and ensure that two components of roughly comparable number of ``old'' and ``young'' stars are present in each system. Our analysis will first focus on these systems to identify the mechanism that leads to the formation of radial 
gradients before considering the full population of simulated dwarfs.
%%

%%%%%%%%%%%%%%
%\begin{table}
%\caption{Adopted values of the parameters of the~\citep{Springel2003} multiphase model for star formation and feedback.}
%\label{TabSFPar}
%\begin{center}
%\begin{tabular}{@{}cccccc}
%\hline
%$\beta$ & $\mathrm{T_{SN}}$ & $\mathrm{\rho_{th}}$ & $\mathrm{t_*}$ & %$\mathrm{T_{c}}$ & $\mathrm{A_0}$  \\
%\hline
%0.1 & $10^8 \ K$ & $\mathrm{0.12 \ cm^{-3}}$ & $\mathrm{3 \ Gyr}$ & $\mathrm{2000 \ %K}$ & 1000 \\
%\end{tabular}
%\end{center}
%\end{table}
%%%%%%%%%%%%%%%%%

\section{Results}
\label{SecResults}

Six simulated dwarfs satisfy the SFH criteria laid out in the previous section (all of them ``isolated''), and we will refer to them hereafter as ``two-component'' dwarfs. As discussed by \citet{Benitez-Llambay2014} the two distinct components form as a result of a merger between two systems whose masses were very close to the hydrogen-cooling limit at the time of cosmic reionization, $z_{\rm reion}$. They are able to form some stars before reionization but, because of their weak potential wells, star formation activity declines soon thereafter as the combined effects of reionization and feedback heat the remaining gas. Not all the gas is lost, however; some remains attached to the galaxy forming a tenuous halo unable to cool and condense to form stars.  This gas serves as fuel for the second episode of star formation, triggered by a subsequent merger between two such systems. The net result is a galaxy with two distinct stellar components clearly separated in age. We proceed below to analyse in some detail 
the kinematics, spatial distribution, and metal content of these two components.

\begin{figure*}
 \begin{center}
 \includegraphics[scale=0.4]{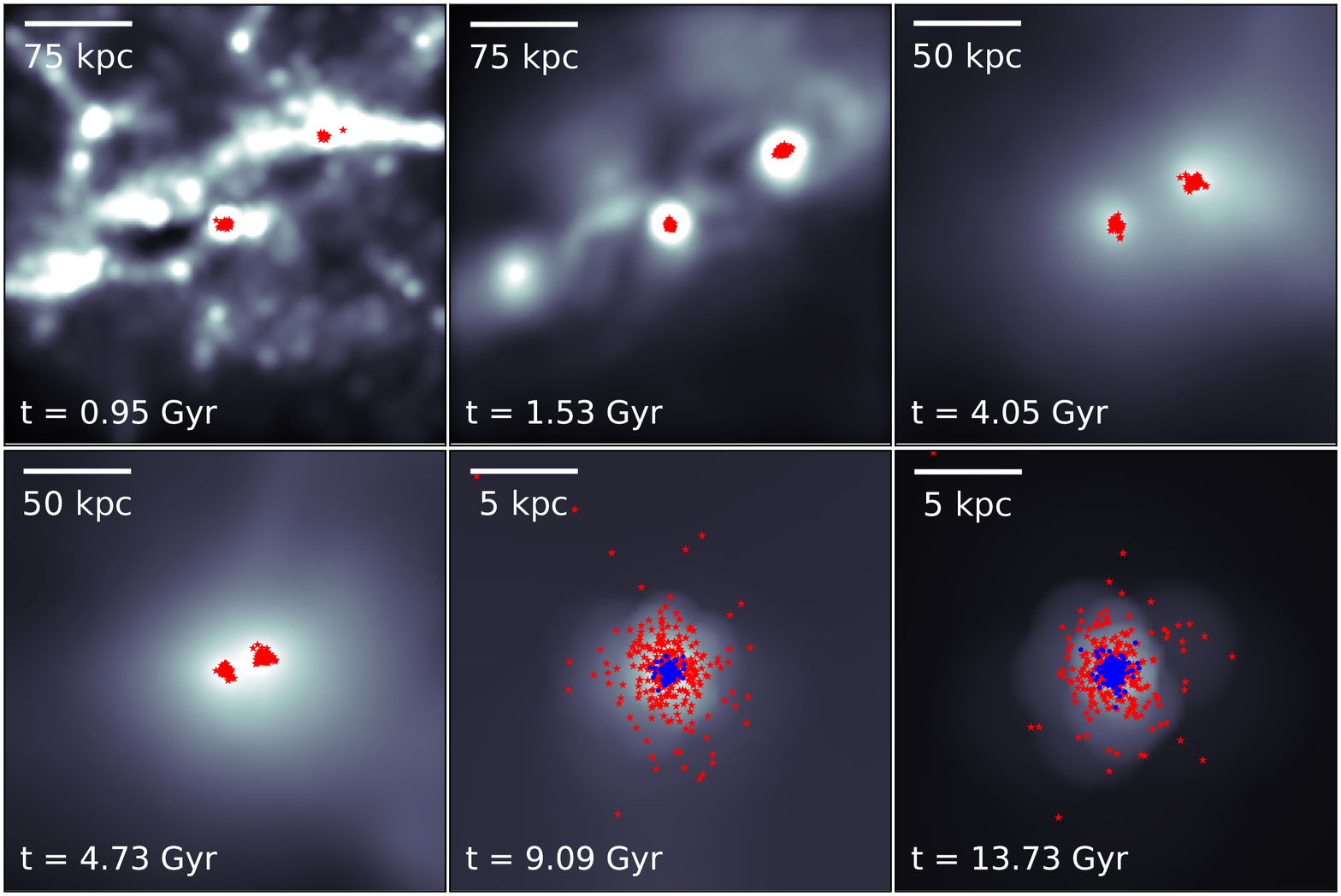}
 \caption{Projected gas and stellar distribution of the ``D4'' dwarf, shown as a function of time. Different shades of grey indicate gas density; red starred symbols denote stars that belong to the ``old'' stellar component at $z=0$ (i.e., those with formation times $t_{\rm form}<4.5 $ Gyr). Blue dots indicate the same, but for ``young'' stars (i.e., $t_{\rm form}\ge 4.5 $ Gyr, see Fig.~\ref{FigAgeMet}). Note that the old stellar population is assembled through a sequence of mergers, which culminate in a rather major event at $t\sim 5$ Gyrs that prompts the in situ formation of the younger component.} \label{FigDwarfEvol}
\end{center}
 \end{figure*}

\subsection{Age and metallicity distributions}

Figure~\ref{FigAgeMet} shows the distribution of formation times and metallicity for the six dwarfs. Metallicities are estimated from the total mass of elements heavier than He and scaled to solar values assuming a solar metallicity of $Z_\odot=0.02$. The two distinct age components used to identify these dwarfs are clearly seen in each panel; the first generation of stars (red solid line) ends abruptly at $t \sim 2$-$3$ Gyr, shortly after reionization. The second episode of star formation (blue dashed line) is more extended, starting at $t\sim 6$ Gyr and lasting in some cases until very recent times.

Note that the two populations differ not only in age (by construction) but also in metallicity; the younger population (in blue dashed line) is systematically more metal rich than the older one (in red solid line), although the metallicity distributions of both are fairly broad and show in some cases significant overlap. Although the distinction between components in metallicity is less clear, for analysis purposes we shall distinguish between a ``metal-poor'' ($[Z/Z_\odot]<-1$) and a ``metal-rich'' ($[Z/Z_\odot]>-1$) population. 

The reason for the well defined trend between age and metallicity is that, although the formation of the younger population is triggered by a gas-rich merger, much of the gas that fuels it had been enriched by the old stellar population. Although this enriched gas was not forming stars after $t\sim 2$ Gyr, it was not completely lost by the galaxy, and it contributes substantially to the second generation of stars when it is brought back to the centre during the merger. 

The only exception is the dwarf labelled ``D6'', where old stars were able to form in only one of the progenitors (the other one was well below the hydrogen-cooling limit at $z_{\rm reion}$). Much of the gas in this case is then relatively unprocessed material, and results in a second episode of star formation that differs only slightly from the first one in terms of metallicity.

\subsection{Spatial distribution}

We now explore the spatial distribution of the two stellar components. Figure~\ref{FigMassProf} shows the projected stellar density profile of one of the simulated dwarfs (labelled as ``D4'' in Figure~\ref{FigAgeMet}). Black circles show to profile of {\it all} stars in the range $0.1 < R/{\rm kpc} < 5$, whereas coloured symbols (red triangles for old/metal-poor, blue squares for young/metal-rich) show the profiles of each of the two components, respectively. Lines indicate King-model fits to each profile ~\citep{King1962}. 

It is clear from Fig.~\ref{FigMassProf} that the stellar mass profile of ``D4'' is well approximated by a King model, and that the young/metal-rich component of ``D4'' is more centrally concentrated than the older, metal-poor component. Indeed, the half-mass radius of the young component is nearly three times smaller than that of the old stars (half-mass radii of each component are shown by arrows in Fig.~\ref{FigMassProf}).  The same segregation is seen after splitting the system by metallicity (right-hand panel) rather than by age, although in that case the distinction is blurred somewhat. 

The spatial segregation noted above applies not only to ``D4'' but to all six dwarfs. We indicate this by the downward-pointing arrows in Fig.~\ref{FigMassProf}, which indicate the half-mass radius of each component for all six simulated dwarfs in our sample. The younger/metal-rich component is always more centrally concentrated than the older one. The degree of segregation is comparable to that observed for the Fornax dSph: the arrows labelled ``Fornax'' in the left-hand panel of Fig.~\ref{FigMassProf} show the ``half-number'' radii quoted by \citet{Battaglia2006} for the intermediate-age and old stellar population in their study.

\subsection{Origin of the spatial segregation}

Why is the younger/metal-rich population always more concentrated? We have traced this to the mode of assembly of each component: the old component is assembled in all cases through ``major'' mergers in which the halo typically more than doubles its mass, whereas the young component is generally formed ``in situ''. This is shown in Fig.~\ref{FigAgeDist}, where we plot the time of formation of each star as a function of the distance from the centre of the main progenitor {\it at the time of formation}. As is clear from this figure, a significant fraction of the old stellar component in all dwarfs formed in a different progenitor, about $70$-$100$ kpc (physical) away from the main system. 

These progenitors merge quickly, in a sequence of events that culminates with the late merger that leads to the formation of the younger component. The series of mergers may be seen in Fig.~\ref{FigDwarfEvol}, which shows several stages of the evolution of the ``D4'' simulated dwarf. Each panel is centred on the most massive progenitor and the grey scale is used to indicate local gas density. Red starred symbols are used to indicate the location of stellar particles that belong at $z=0$ to the ``old'' population (i.e., $t_{\rm form}\simlt 2.5$ Gyr) and that have formed by the current time. Blue dots show the location of stars belonging to the ``young'' stellar population.  

This figure shows clearly how the old component is assembled mainly through a merger which occurs at $t\sim 5$ Gyrs. Note that each of the two merging objects is surrounded by gas but are not forming stars just before the merger. This is the gas that fuels the in-situ formation of the younger stellar component, as the merger compresses the gas and allows it to cool.

\begin{figure}
 \begin{center}
 \includegraphics[scale=0.5]{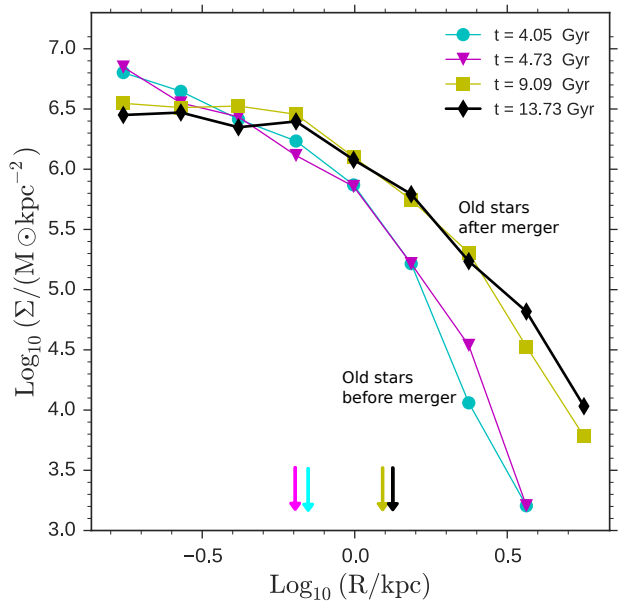}
 \caption{Surface density profile of the old stellar population in the main progenitor of the ``D4'' simulated dwarf, at different times. The time sequence is chosen to illustrate the changes in the density profile of the old stars that occur as a result of the merger shown in Fig.~\ref{FigDwarfEvol} at $t\sim 5$ Gyr. Before the merger the system is not forming stars, and the profiles are stable. After the merger, the old stellar component increases in mass and radius, as may be seen by comparing the arrows, which indicate the half-mass radius at each time. The merger is clearly responsible for distending the old stellar component, lowering its density and leading to a more extended population that the young stars, which form largely in situ.} \label{FigDensProfEvol}
 \end{center}
\end{figure}

The merger also causes the old stellar population of the main progenitor to roughly double in mass and to increase its size by roughly $80\%$, decreasing its density relative to the newly-formed stellar component. We show this in Fig.~\ref{FigDensProfEvol}, which shows the projected stellar density profile of the old component in the main progenitor before and after the merger. Before $t\sim 5$ Gyr, the galaxy was not forming stars and the stellar density profile was stable and quite centrally concentrated. After the merger (the last two times shown in the figure by the squares and diamond-like symbols), the profile of old stars (which now includes the old component of the merging partner) has again settled into equilibrium, but at much lower central density and with a larger effective radius. The merger is thus clearly responsible for the spatial segregation between components.

\subsection{Kinematics}

The structural parameters of each dwarf, which are listed in Table~\ref{TabGxProp} show that, as expected from their spatial segregation, the velocity dispersion of the younger component is always lower than that of its surrounding older population. Interestingly, the magnitude of the difference (on average $20\%$) is comparable to the difference that \cite{Battaglia2006} report in Fornax. Both populations are mildly radially anisotropic, and in only one case (the young component of dwarf ``D1'') tangential motions dominate. This is also the dwarf where stars have formed most recently (see Fig.~\ref{FigAgeMet}), and where the youngest stars show some sign of rotational support.

At $z=0$ ordered rotation is subdominant in both stellar components, but the gas component seems to be rotationally supported in all galaxies. Typically, $70 \pm 15 \%$ of the gas particles are in co-rotating orbits. Although we have attempted to trace in detail the evolution of the rotational support of each stellar component the results were inconclusive, mainly due to the noise in the measurements imposed by the limited number of star particles in each system (of order a few hundred), and by the fact that these galaxies are not well resolved spatially; indeed, the average half-mass radius of the young stellar component is only of order $\sim 4$ times the gravitational softening at redshift $z=0$. We therefore defer a more detailed study of the rotational properties of these galaxies to future simulations with better mass and spatial resolution.

\subsection{Metallicity gradients}

\begin{figure}
 \begin{center}
 \includegraphics[scale=0.45]{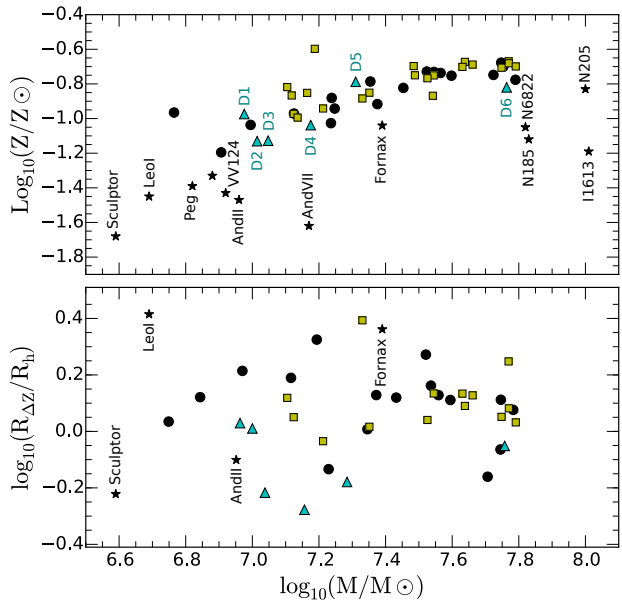}
 \caption{Top panel shows the mass-metallicity relation for Local Group dwarf galaxies (black stars) and for CLUES dwarfs in the stellar mass range $10^{6.5} < M/M_{\odot} < 10^{8}$. Cyan upward-pointing triangles show the six ``two-component'' dwarfs of our previous analysis. Black circles show the remaining isolated CLUES dwarfs in the same stellar mass range, while yellow squares indicate satellite systems. The bottom panel shows the radius where the mean metallicity of a dwarf drops by 0.2 dex from its central value (see Figure~\ref{FigMetslope}), normalized to the stellar half-mass radius, as a function of stellar mass. Data taken from~\citet{Leaman2013},
 except for AndII, AndVII and NGC 185, which were taken from~\citet{Ho2015}. Half-mass radii were taken from~\citet{McConnachie2012}.} \label{FigMetRel}
 \end{center}
\end{figure}

% new
We explore next the metallicity gradients of two-component systems and compare them with the rest of the simulated sample, which we call, for simplicity, ``single-component'' dwarfs. We first note that two-component dwarfs do not differ from other simulated dwarfs in their average metallicity, at fixed stellar mass. We show this in the top panel of Fig.~\ref{FigMetRel}, where we plot the mean metallicity of all simulated dwarfs as a function of their stellar mass. Upward cyan triangles correspond to two-component simulated dwarfs; black circles are ``single-component'' systems, and yellow squares correspond to satellites. The metallicity of simulated dwarfs compares favourably with those of nearby dwarfs of similar mass (starred symbols).

As expected from our earlier discussion, two-component dwarfs show steeper gradients than ``single-component'' dwarfs. This is shown in Fig.~\ref{FigMetslope}, where we plot the mean metallicity profile, normalized to the central metallicity value and expressed in units of the stellar half-mass radius.  Cyan-solid lines in Fig.~\ref{FigMetslope} indicate the metallicity profile averaged over the full sample of simulated ``two-component'' dwarfs; the black-dashed line, on the other hand, correspond to the remaining dwarfs. We quantify the metallicity gradient using the ratio between the stellar half-mass radius, $R_{\rm h}$, and the radius, $R_{\Delta Z}$, where the metallicity drops by $0.2$ dex from its central value. The larger the ratio, $R_{\Delta Z}/R_{\rm h}$, the shallower the metallicity gradient: two-component dwarfs clearly have steeper gradients than the rest.

Interestingly, the two nearby dwarfs where the presence of distinct components is arguably most obvious (Sculptor and And II) have the steepest gradients as well, according to this metric (we note, however, that these are satellite galaxies, so the comparison is not straightforward). We show this in the bottom panel of Fig.~\ref{FigMetRel}, where $R_{\Delta Z}/R_{h}$ is plotted as a function of the stellar mass of each simulated dwarf. Despite the admittedly small sample, it is tempting to ascribe such gradients to the role of mergers during their formation.

\subsection{Mergers and metallicity gradients in dwarfs}

\begin{figure}
 \begin{center}
 \includegraphics[scale=0.5]{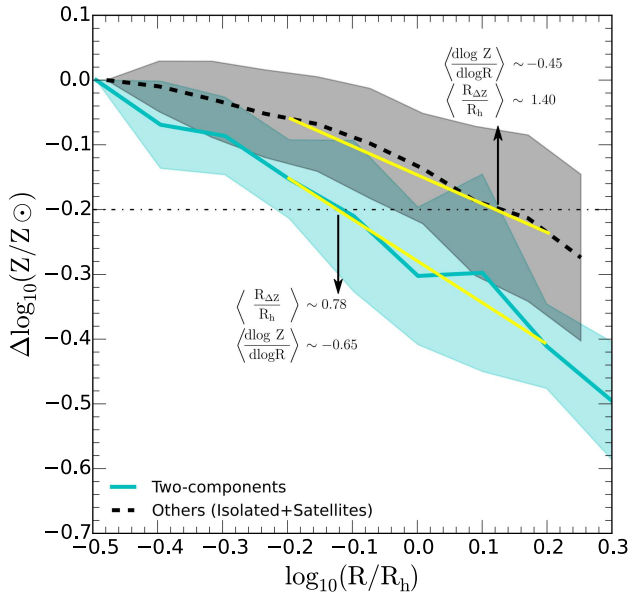}
 \caption{Average metallicity profile normalized to the central metallicity value as a function of radius in units of the ``half-mass'' radius. Cyan-solid and black-dashed lines show the average profile of the six simulated ``two-component'' galaxies and of the remaining dwarfs (isolated and satellites) in the stellar mass range ($10^{6.5} < M/M_{\odot} < 10^{8}$), respectively. Shaded areas show the typical {\it rms} per bin. ``Two-component'' dwarfs show, on average, a steeper metallicity gradient compared to those where a single stellar population prevails. We quantify the steepness of the metallicity gradient by computing the $(R_{\Delta Z}/R_h)$ ratio, where $R_{\Delta Z}$ is the radius where the metallicity drops by $0.2$ dex from its central value. Yellow solid straight lines indicate the metallicity slope corresponding to the given $R_{\Delta Z}/R_h$ value.}
 \label{FigMetslope}
 \end{center}
\end{figure}

\begin{figure}
 \begin{center}
 \includegraphics[scale=0.65]{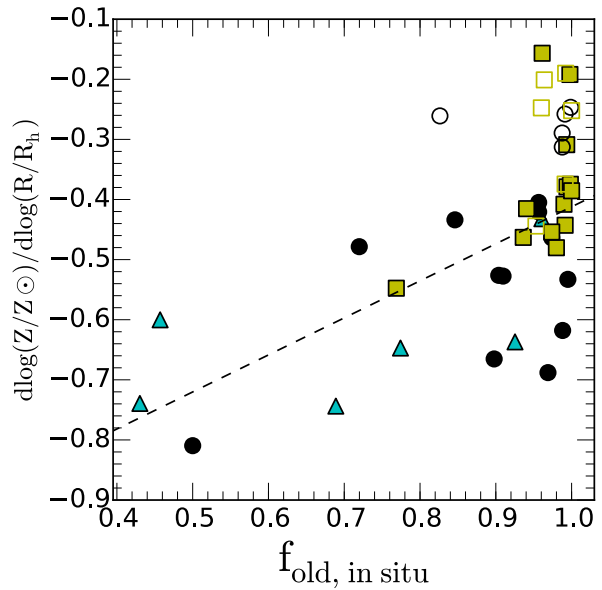}
 \caption{Logarithm slope of the metallicity gradient of dwarf galaxies as a function of the fraction of old stars formed in situ. The six upward cyan triangles correspond to the ``two-component'' systems. Circles are galaxies dominated by a single old stellar population and yellow squares show the satellite dwarfs in the given stellar mass range ($10^{6.5} < M/M_{\odot} < 10^{8}$). This figure confirms that galaxies that have accreted a significant fraction of old stars develop steeper metallicity gradients, and the trend is independent on whether a dwarf is satellite or isolated. Empty and filled symbols show dwarfs that have formed more and less than 90\% in the first $4 \ Gyr$ respectively. Dashed line is the best linear fit to the data.}
 \label{FigFracAcc}
 \end{center}
\end{figure}

Are gradients also related to merger activity in ``single component'' systems; i.e., dwarfs with less obviously distinct components or a more continuous SFH? We explore this in Fig.~\ref{FigFracAcc}, where we show the logarithmic slope of the metallicity gradient measured at $R_{\rm h}$ as a function of the fraction of old stars formed in the main progenitor, or ``in situ'', $f_{\rm old, in situ}$. Systems where $f_{\rm old, in situ}$ is low (i.e., where mergers are important in the assembly of the old stellar component) have in general strong gradients\footnote{We use in Fig.~\ref{FigFracAcc} the logarithmic slope of the metallicity radius relation measured at the stellar half-mass radius as a measure of the gradient. This is equivalent to the $R_{\Delta Z}$ parameter introduced in Fig.~\ref{FigMetslope}. Although the latter is easier to estimate from observations, it becomes ill-defined for very weak gradients.}. This is not only true for ``two-component'' dwarfs (cyan triangles), but also for systems with 
more continuous SFHs, and even for satellites (although only one satellite in our sample has $f_{\rm old, in situ}<0.9$; see yellow squares). 

At the same time, it is also clear from Fig.~\ref{FigFracAcc} that steep gradients are present even in dwarfs where mergers have not played a substantial role in the assembly of the old stellar population (i.e., when $f_{\rm old, in situ}>0.9$). For such systems, a clear distinction may be seen between those that formed essentially all stars in the first $4$ Gyr (i.e., $f_{\rm old}>0.9$, shown as open symbols), which have weak gradients, and those where star formation continued until more recent times (i.e., $f_{\rm old}<0.9$, indicated by filled symbols), whose gradients are stronger.

Fig.~\ref{FigFracAcc} thus shows that systems that formed monolithically are those where gradients are least obvious, whereas most systems with a protracted star formation history generally tend to develop gradients, with mergers accentuating the effect.  This is because most dwarfs undergo some merging activity at some point in their lifetime; those mergers will induce gradients provided they are followed by subsequent star formation. Two conditions therefore seem necessary to establish strong gradients: (i) the occurrence of a merger able to disperse the (typically metal-poor) stars formed prior to the event, and (ii) the subsequent, post-merger formation of a more centrally concentrated in situ stellar population.

Fig.~\ref{FigFracAcc} also shows that simulated satellites (yellow symbols) have in general weaker gradients than isolated objects. This is because the accretion of a dwarf into a massive halo affects the two conditions listed above: ram-pressure stripping robs satellites of the gaseous fuel needed to sustain a protracted star formation history, and dwarf-dwarf mergers become less frequent due to the high orbital velocities of satellites compared to their internal velocity dispersion. Satellites thus form more monolithically than field dwarfs; they experience fewer mergers (i.e., $f_{\rm old, in situ}>0.9$, with a single exception) and they also have little residual star formation at more recent times (the mean satellite $f_{\rm young}$, for example, is $\sim 3 \%$ compared with $\sim 11\%$ for field dwarfs of comparable stellar mass). Our results suggest that strong gradients should be more easily detectable in isolated galaxies than satellites, a prediction that should be testable once reliable gradients 
become available in a statistically-significant sample of both satellite and isolated spheroidal galaxies.

\section{Summary and Conclusions}
\label{SecConc}

We have studied the ``outside-in`` formation of dwarf spheroidals using a cosmological hydrodynamical simulation of the Local Group. We focus on systems chosen to have formed large fractions of their stars at early times ($t_{\rm form}<4$ Gyr) and in the recent past ($t_{\rm form}>8$ Gyr) so that they contain, by construction, two stellar components of distinct age.  These ``two-component'' systems are expected to develop strong radial gradients in age and metallicity and were examined to search for clues to the origin of such gradients.

The two-component simulated dwarfs form in low-mass halos that were close to the hydrogen-cooling limit at the time of reionization and that undergo a major merger at late times. Because of their low mass they are able to form a limited number of stars before reionization. The feedback energy from those stars, together with cosmic reionization, are able to heat the gas and bring the early episode of star formation to a temporary close. This process, however, does not remove the gas completely from the dwarf, but leaves some in a tenuous halo around each dwarf.  This gas is compressed and allowed to cool and condense during a subsequent merger, fuelling a second major episode of star formation.  The merger pumps energy into the old/metal-poor stellar component of the galaxy, which expands to lower densities and is at late times seen to surround the younger, more metal-rich population, which forms largely in situ. This process leads to the outside-in formation of two spatially-segregated stellar systems with 
distinct kinematics and metallicity.

These general trends agree with observations of dwarf spheroidals where multiple components have been identified, such as Sculptor or And II, including their relatively steep metallicity gradients. Our results thus suggest a natural explanation for the observed differences between components and imply that major mergers have likely played an important role in the formation of such systems. This inference is supported by the discovery of convincing evidence for external accretion in And II \citep{Amorisco2014}, and by the results of \citet{Deason2014}, who argue, on more theoretical grounds, that relatively few isolated Local Group dwarfs could have avoided a major merger in their formation history.  

The same merger mechanism can also induce gradients in systems with more continuous star formation histories, even if in those cases the gradients found are less pronounced. We find that two conditions should be satisfied for steep gradients to develop: (i) some merger event must occur to disperse the old, metal-poor stellar population, and (ii) mergers must be followed by residual in situ star formation.

Following this line of thought, one is tempted to speculate that mergers are the root cause of the metallicity and age gradients seen in a large fraction of dwarf galaxies, spheroidal and irregular alike \citep[see, e.g.,][for some recent work on gradients in dSph and dIrr galaxies]{Harbeck2001,Bernard2008,Hidalgo2013,Bellazzini2014}. Seen from that perspective, two-component systems are just extreme examples where there is a hiatus in the star formation history that is rekindled by a major merger. Minor mergers, on the other hand, might be responsible for setting up gradients in systems where the dichotomy is less obvious.  Confirming this ansatz would require identifying the telltale signatures of past mergers in isolated dwarfs, such as low surface brightness ``shell-like'' structures in the outskirts, or the presence of kinematic anomalies that might be traced to external events. Until such evidence is uncovered and understood it might be too early to conclude that mergers have played a substantial role 
in the formation of most isolated dwarfs. Our results, however, argue that this is the case in at least some.

\section{Acknowledgements}

We thank the anonymous referee for valuable comments that helped to improve the paper. We also thank to Alan McConnachie, Amina Helmi and Laura Sales for useful discussions. This research also was supported by the National Science Foundation under Grant No. PHY11-25915 and the hospitality of the Kavli Institute for Theoretical Physics at UC Santa Barbara. The simulations were performed at Leibniz Rechenzentrum Munich (LRZ) and at Barcelona Supercomputing Center (BSC). Our collaboration has been supported by DFG grants GO 563/21-1 and GO 563/24-1 as well as by CONICET. ABL, JFN and MGA aknowledge support from ANPCyT grant PICT2012- 1137. GY acknowledges support from the Spanish MINECO under research grants AYA2012-31101 and  FPA2012-34694. YH has been partially supported by the Israel Science Foundation (1013/12).

This paper has benefited from Ipython~\citep{PER-GRA:2007} and matplotlib~\citep{Hunter:2007}.

\begin{table*}
\caption{Table listing the main parameters of the cosmological simulation. We use the usual nomenclature to list the cosmological parameters \citep{Spergel2007}. The last four columns list, respectively, the particle mass of the dark matter, gas, and stellar components, as well as the (comoving) gravitational softening of the high-resolution region.}
\begin{tabular}{@{}llllllllll}
\hline
\hline
$\Omega_M$ & $\Omega_b$ & $\Omega_{\Lambda}$ & $h$ & $\sigma_8$ & $n$ & ${\rm m_{drk}}$ & ${\rm m_{gas}}$ & ${\rm m_{str}}$ & $\epsilon$ \\
0.24 & 0.042 & 0.76 & 0.73 & 0.75 & 0.95 & $2.87 \times 10^5 M_{\odot}$ & $6.06 \times 10^4 M_{\odot}$ & $3.03 \times 10^4 M_{\odot}$ & 137 pc \\
\hline
\end{tabular}
\label{TabSimParam}
\end{table*}

\begin{table*}%\adjustbox{max width=\columnwidth}{
{\caption{Properties of the ``two-component'' galaxies: total stellar mass ($M_{\rm gal}$); fraction of stars in each component ($f_{\rm old}$, $f_{\rm young}$); stellar velocity dispersion ($\sigma$); stellar radial velocity dispersion ($\sigma_r$); anisotropy parameter $\beta=1- \sigma_t^2/2\sigma_r^2$; mean stellar metallicity ($Z$) in log solar units; and the stellar half mass radii ($R_h$).  Values enclosed in parentheses or square brackets are computed considering the ``old'' and ``young'' components, respectively. }
\begin{tabular}{cccccccccc}
\hline
ID & $M_{200}$ & $M_{\rm gal}$ & $(f_{\rm old}$)[$f_{\rm young}$] & $\sigma$/(km/s) & $\sigma_r$/(km/s) & $\beta$ & $[Z/Z_{\odot}]$ & $R_h/{\rm kpc}$ & \\
   &      [$10^9 \, M_{\odot}$]               &    [$10^7 \, M_{\odot}$]       & & total(old)[young] & total(old)[young] & total(old)[young] & total(old)[young] & total(old)[young] & \\
\hline
D1 & 3.74 & $0.92$ & (0.15)[0.85] & 19.97(26.00)[18.74] & 11.08(15.18)[10.24] & -0.13(0.03)[-0.17] & -1.10(-1.60)[-0.65]& 0.49(1.40)[0.42]\\
D2 & 2.95 & $1.00$ & (0.75)[0.25] & 22.62(26.35)[20.45] & 13.24(13.53)[12.17] & 0.04(0.01)[0.09] & -1.17(-1.54)[-0.65]& 1.00(1.22)[0.59]\\
D3 & 4.32 & $1.10$ & (0.70)[0.30] & 23.09(24.48)[19.79] & 13.98(14.50)[12.82] & 0.14(0.08)[0.3] & -1.19(-1.55)[-0.65]& 0.87(1.05)[0.67]\\
D4 & 3.96 & $1.4$ & (0.61)[0.39] & 25.02(25.88)[23.55] & 16.46(17.35)[15.00] & 0.34(0.39)[0.27] & -1.12(-1.56)[-0.63]& 0.85(1.30)[0.49]\\
D5 & 3.94 & $1.9$ & (0.25)[0.75] & 23.82(29.09)[21.89] & 14.71(19.17)[12.98]& 0.19(0.35)[0.08] & -0.90(-1.58)[-0.54]& 0.77(1.66)[0.63]\\
D6 & 7.00 & $5.7$ & (0.47)[0.53] & 27.12(28.24)[26.35] & 16.71(17.12)[16.43] & 0.18(0.14)[0.21] & -0.92(-1.51)[-0.58] & 0.91(1.14)[0.75]\\
\end{tabular}
\label{TabGxProp}
}
\end{table*}

\bibliographystyle{mn2e}
\bibliography{my_biblio}
\end{document}